\documentclass[aps,prb,twocolumn,secnumarabic,balancelastpage,amsmath,amssymb,
floatfix,showpacs,longbibliography,reprint,superscriptaddress]{revtex4-1}

\usepackage{amsmath, amssymb}
\usepackage{empheq}
\usepackage{braket}
\usepackage{natbib}

\newcommand{\cuo}{Cu$_{\text{2}}$O} 
\newcommand{\sy}[2][+]{\Gamma^{#1}_{#2}} 

\newcommand{\bk}{\mathbf{k}}
\newcommand{\bK}{\mathbf{K}}

\newcommand{\br}{\mathbf{r}}

\newcommand{\bR}{\mathbf{R}}

\newcommand{\basev}[2]{\Ket{\begin{smallmatrix}
                           {\scriptstyle #1}\\{\scriptstyle #2}
                          \end{smallmatrix}}}

\begin{document}
                          
\title{Interactions between Rydberg excitons in Cu$_2$O}

\author{Valentin Walther}
\affiliation{Department of Physics and Astronomy, Aarhus University, Ny 
Munkegade 120, DK 8000 Aarhus C, Denmark} 
\author{Sjard Ole Kr\"uger}
\affiliation{Institut f\"ur Physik, Universit\"at Rostock, 
Albert-Einstein-Stra{\ss}e 23, D-18059 Rostock, Germany}
\author{Stefan Scheel}
\affiliation{Institut f\"ur Physik, Universit\"at Rostock, 
Albert-Einstein-Stra{\ss}e 23, D-18059 Rostock, Germany}
\author{Thomas Pohl}
\affiliation{Department of Physics and Astronomy, Aarhus University, Ny 
Munkegade 120, DK 8000 Aarhus C, Denmark}

\begin{abstract}
Highly-excited states of excitons in cuprous oxide have recently been observed 
at a record quantum number of up to $n=25$. Here, we evaluate the long-range 
interactions between pairs of Rydberg excitons in Cu$_2$O, which are due to 
direct Coulomb forces rather than short-range collisions typically considered 
for ground state excitons. A full numerical analysis is supplemented by the van 
der Waals asymptotics at large exciton separations, including the angular 
dependence of the potential surfaces.
\end{abstract}

\maketitle

\section{Introduction}
Excitons play an important role for the optical properties of many 
semiconductors. Composed of an electron and a hole bound by their Coulomb 
attraction, excitons may be considered as artificial atoms that feature a series 
of energy levels very similar to that of simple one-electron atoms. 
Their
relatively low exciton binding energies
combined with additional effects such as phonon coupling \cite{schweiner2016} or crystal 
inhomogeneities, however, render the observation of excited exciton states 
inherently difficult. Cuprous oxide (Cu$_2$O) stands out in this respect, as it 
features a comparably large Rydberg energy of $\sim86$ meV, which together with 
the narrow absorption lines provides well-suited conditions for exciting 
excitonic Rydberg states. In fact, recent measurements on Cu$_2$O semiconductors 
\cite{kazimierczuk2014giant} have demonstrated the preparation of highly-excited 
Rydberg excitons with record-breaking principal quantum numbers of up to 
$n=25$. This discovery has sparked renewed theoretical and experimental interest in the field of excitons, ranging from excitonic spectra in magnetic \cite{zielinska2017} and electric \cite{heckoetter2017, zielinska2016} fields as well as non-atomic scaling laws \cite{heckoetter20172} to the breaking of all antiunitary symmetries \cite{schweiner2017} and the onset of quantum chaos \cite{assmann2016}. 

A further particular appeal of such Rydberg states stems from their strong mutual 
interactions, which, as demonstrated for cold atomic systems \cite{saffman2010}, 
can lead to enhanced optical nonlinearities of the material 
\cite{walther2018giant}. In contrast to ground-state excitons whose low-energy 
interactions can often be described in terms of zero-range collisions 
\cite{cuiti1998,tassone1999}, the interaction between Rydberg excitons can 
become important already on much larger length scales and lead to an exciton 
blockade \cite{kazimierczuk2014giant} that prevents the optical excitation of 
two excitons within typical distances of several $\mu$m. Under such conditions  
the relevant interactions are no longer dominated by exchange effects 
\cite{cuiti1998} but are determined by direct Coulomb interactions between the 
excitons. 

In this work, we determine the interaction between Rydberg excitons in Cu$_2$O. 
Our calculations account for the dipole-dipole coupling between energetically 
close exciton pair states that is induced by their direct Coulomb interaction 
and dominates the overall interaction at the large distances relevant under 
experimental conditions \cite{kazimierczuk2014giant} of Rydberg exciton 
blockade. Asymptotically, the interaction is of van der Waals type with a van 
der Waals coefficient that is found to follow a simple scaling law which was 
previously used for Rydberg state interactions of alkaline \cite{singer2005} and 
alkaline earth atoms \cite{mukherjee2011}. From our calculations, we determine 
the corresponding scaling coefficients, providing easy access to precise values 
of Rydberg-exciton van der Waals coefficients in Cu$_2$O for future studies of 
many-body effects or nonlinear optical phenomena due to interactions between 
highly-excited excitons.  

The article is organized as follows. After outlining the determination of 
Rydberg exciton wave functions from the semiconductor band structure in 
Sec.~(\ref{sec:single}), we describe our calculations of the direct Coulomb pair 
interaction in Sec.~(\ref{sec:interaction}). The obtained potential energy 
curves are discussed in Sec.~(\ref{sec:vdW}), where we present the 
perturbative calculation of the van der Waals interactions and summarize our 
results for the van Waals coefficients for a broad range different excitonic 
Rydberg states. Finally, implications, limitations and potential applications of 
the results are discussed in Sec.~(\ref{sec:discussion}).

\section{Single Exciton states} \label{sec:single}

Bulk {\cuo} is a semiconductor with a cubic crystal structure of the point 
group $O_h$ and a direct band gap of $E_g=2.17208~\text{eV}$ at the center of 
its Brillouin zone \cite{kazimierczuk2014giant}. Without spin, the uppermost 
valence band has $\sy{5}$ symmetry, which is split into an upper $\sy{7}$- and a 
lower $\sy{8}$-band by the spin-orbit interaction. These two bands are separated 
by a corresponding spin-orbit splitting of $\Delta=131~\text{meV}$ and can be 
described by an effective band Hamiltonian derived in Ref.~\cite{suzuki1974}. 

Together with the lowest $\sy{6}$ conduction band, these valence bands form two 
excitonic series: the so-called yellow ($\sy{6}\otimes\sy{7}$) and green 
($\sy{6}\otimes\sy{8}$) series. The optical transition from the excitonic vacuum 
to the $s$-excitons is dipole forbidden for both series due to the positive 
parity of both the conduction and the valence band. The series of interest to 
this work is the yellow series whose $p$-exciton resonances are located below 
the band gap and have been observed experimentally \cite{kazimierczuk2014giant}.

We determine the exciton binding energies, $E_{{\bf K},nl}$, and wave 
functions, $\tilde{\Psi}({\bf K},{\bf k})$, from the nonparabolic momentum-space 
Wannier equation
\begin{equation}
\begin{aligned}
\left[\frac{\hbar^2\left(\alpha\bK + \bk\right)^2}{2m_e} + 
T_h\left(\left|\beta\bK - \bk\right|^2\right) \right]~\tilde{\Psi}(\bK,\bk) \\
+ \frac{e^2}{8\pi^3\varepsilon_0\varepsilon_r}\int 
d^3\bk'~\frac{\tilde{\Psi}(\bK,\bk')}{|\bk-\bk'|^2}  = E~\tilde{\Psi}(\bK,\bk) 
\label{eq:nonpara-wannier}
\end{aligned}
\end{equation}
as described in Ref.~\cite{schoene2016}. Here, the hole's dispersion $T_h$ is 
obtained from an angular average over the hole dispersion derived from the 
valence band Hamiltonian of Ref.~\cite{suzuki1974} and $\varepsilon_0$ denotes 
the vacuum permittivity, while $\varepsilon_r \approx 7.5$ is the static 
relative permittivity of {\cuo}.
Furthermore, $\bk$ and $\bK$ denote the relative and center-of-mass (COM) 
momentum of the electron-hole pair, respectively, and $\alpha = m_e/M$ and 
$\beta = m_h/M$ denote the mass of the electron ($m_e$) and the hole ($m_h$) in 
units of the total exciton mass $M=m_e + m_h$.

\begin{figure}[t!]
 \begin{center}
  \includegraphics[height=.2\textwidth]{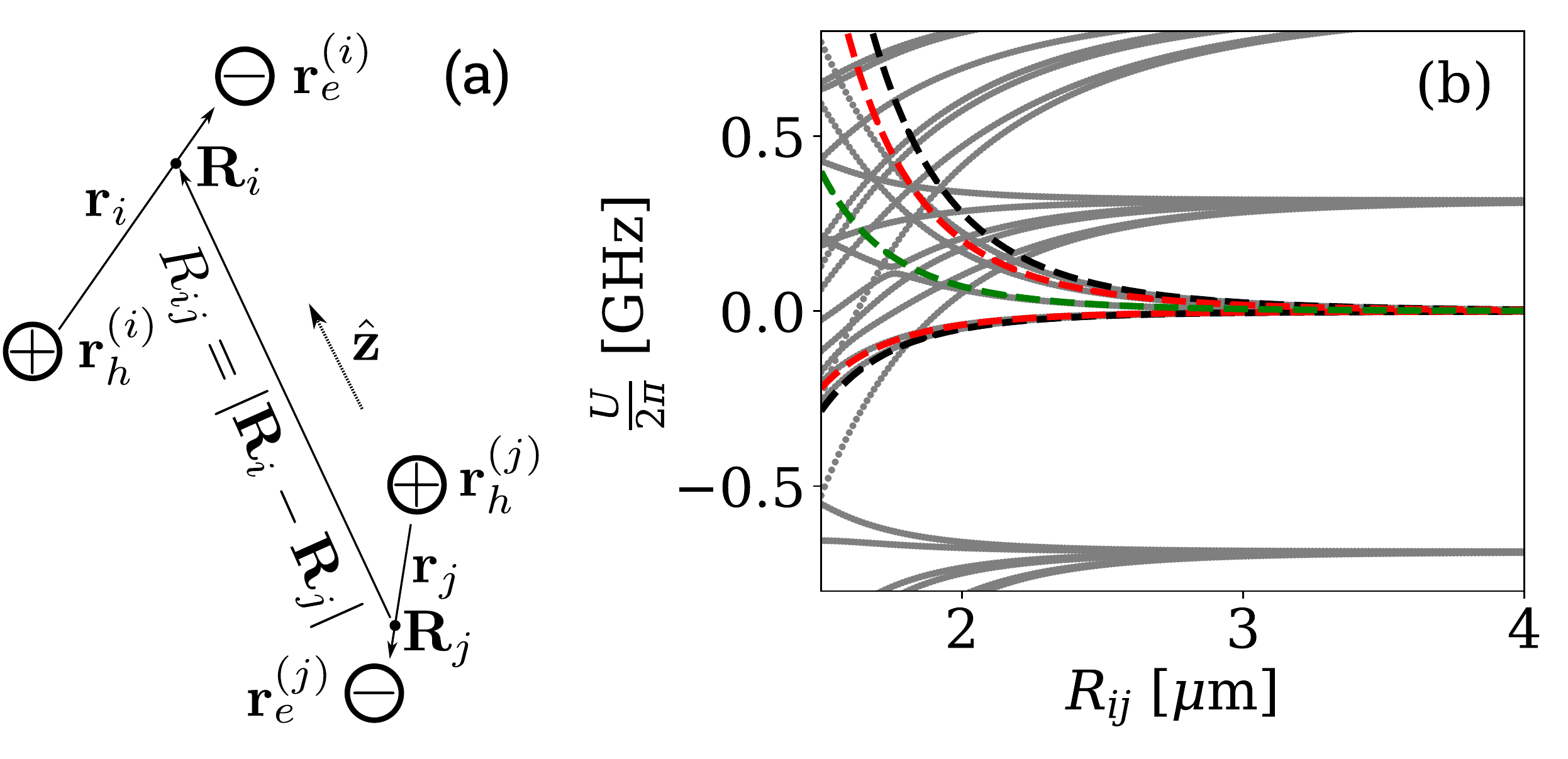}
 \end{center}
 \caption{a) Sketch of a pair of excitons $i$ and $j$, consisting of electrons 
at $\mathbf{r}_e^{(i),(j)}$ and holes at $\mathbf{r}_h^{(i),(j)}$. The center of 
mass coordinates are indicated by $\mathbf{R}_{i,j}$, the exciton separation by 
$R_{ij}$. The coordinate system is aligned with $\hat{\mathbf{z}}$. b) 
Potential energy surfaces centered around $n=15p$ with corresponding 
van-der-Waals curves. The thin lines are obtained from a numerical 
diagonalization, whereas the colored lines show the asymptotic results for the 
different families of $M$ with $|M|=0$ (black), $|M|=1$ (red), $|M|=2$ (green). 
}
\label{fig:C6fit}
\end{figure}

The nonparabolicity of the hole dispersion $T_h$ plays an important role for 
the bound state properties and yields the leading contribution to the excitonic 
quantum defect \cite{schoene2016}. Its effect on the center-of-mass dynamics 
with momentum $K$ can, however, be neglected as long as $K\ll\pi/a_l$, where 
$a_l$ is the lattice constant. This approximation is well justified because the 
momentum of the optical photon that generates the exciton is much smaller than 
$\pi/a_l$.
Therefore, we can separate the relative and COM part of the exciton wave 
function, whose real-space representation can consequently be written as
\begin{equation}\label{eq:exciton}
\Psi_{\bK, nlm}(\bR, \br) = \frac{1}{\sqrt{V}} e^{i\bK\cdot\bR}~ \psi_{nlm}(\br),
\end{equation}
with corresponding energies
\begin{equation}
E_{\bK,nl} = \frac{\hbar^2\bK^2}{2M} + E_{0,nl}.
\end{equation}
Moreover, $\bR = \alpha\br_e + \beta\br_h$ and $\br = \br_e - \br_h$, as 
illustrated Fig.~\ref{fig:C6fit}a, and $\psi_{nlm}(\br)$ denotes the 
bound-state wave function obtained from the extended Wannier equation with the 
standard quantum numbers $n$, $l$ and $m$. 

As we have assumed rotational symmetry and neglected the non-parabolic COM 
dispersion, the excitonic states are degenerate with regard to the magnetic 
quantum number $m$. The anisotropy of the valence band can be included in this 
calculation and would lead to further splitting of states with $l\ge 2$. The 
size of this splitting depends on the momentum-space extension and scales 
roughly with $n^{-3}$.  The same is true for exchange-splitting of the 
S-excitons and both effects are neglected in this work, as they are of minor 
importance to the Rydberg states of interest.

\section{Rydberg exciton interaction potential} \label{sec:interaction}
The pairwise interaction between excitons is given by the sum 
\begin{equation}\label{eq:Vij}
\begin{aligned}
 V^{(ij)} = \frac{e^2}{4\pi\varepsilon_0 \varepsilon_r} \left( 
\frac{1}{|\mathbf{r}_{e}^{(i)} - \mathbf{r}_{e}^{(j)}|} + 
\frac{1}{|\mathbf{r}_{h}^{(i)} - \mathbf{r}_{h}^{(j)}|} \right.\\
 \left. - \frac{1}{|\mathbf{r}_{e}^{(i)} - \mathbf{r}_{h}^{(j)}|} - 
\frac{1}{|\mathbf{r}_{e}^{(i)} - \mathbf{r}_{h}^{(j)}|} \right)
\end{aligned}
\end{equation}
of mutual Coulomb interactions between the electron and hole of one exciton at 
respective positions $\mathbf{r}^{(i)}_{e}$ and $\mathbf{r}^{(i)}_{h}$, 
respectively, and another electron-hole pair at positions 
$\mathbf{r}^{(j)}_{e}$ and $\mathbf{r}^{(j)}_{h}$. Within a multipole expansion, 
the interaction can be rewritten \cite{singer2005} as a series of inverse powers 
of the exciton COM distance $R_{ij} = |\mathbf{R}_{i} - \mathbf{R}_{j}|$
\begin{align}
 V^{(ij)} &=  \frac{e^2}{4\pi\varepsilon_0 \varepsilon_r} \sum_{l,L=1}^{\infty} 
\frac{\mathcal{V}_{lL}(\mathbf{r}_i, \mathbf{r}_{j})}{{R_{ij}}^{l+L+1}} 
\label{eq:multipole_expansion},
\end{align}
where
\begin{align}
\mathcal{ V}_{lL}(\mathbf{r}_i, \mathbf{r}_{j}) &= \frac{(-1)^L 
4\pi}{\sqrt{(2l+1)(2L+1)}} r^l_i r^L_j\\
& \sum_m \sqrt{\binom{l+L}{l+m}\binom{l+L}{L+m}}  Y_{lm} 
(\hat{\mathbf{r}}_{i})Y_{L-m} (\hat{\mathbf{r}}_{j})
\end{align}
and $Y_{lm} $ denotes the spherical harmonics defined with respect to the 
distance vector ${\bf R}_{ij}$. 

While the interaction between ground-state excitons \cite{cuiti1998} can be 
often estimated from first-order perturbation theory, by evaluating Coulomb 
scattering matrix elements based on Hartree-Fock states for pairs of 
interacting excitons, such an approximation \cite{shahnazaryan2016} becomes 
inapplicable for excitonic Rydberg states whose large polarizability 
\cite{walker2008, kazimierczuk2014giant, saffman2010} requires a 
non-perturbative treatment of the Coulomb interactions. In this regime, the 
exciton interaction predominantly stems from the virtual dipole-dipole coupling 
between exciton bound states while exchange effects are negligibly small. This 
is typically the case for exciton distances  \cite{leroy1974}
\begin{align}
R_{ij} \gg 2 \cdot \left(\sqrt{\langle r_i^2}  \rangle+\sqrt{\langle r_j^2 
\rangle}\right).
\end{align}
Note that this condition also ensures convergence of the above multipole 
expansion, Eq.~(\ref{eq:multipole_expansion}), which for sufficiently large 
distances is predominantly determined by the dipole-dipole contribution $l=L=1$, 
such that
\begin{equation}\label{eq:V_dd}
\begin{aligned}
 V^{(ij)} \approx \frac{e^2}{4\pi\varepsilon_0 \varepsilon_r} \left( \frac{r_i 
r_j}{R_{ij}^3} - \frac{3(\mathbf{r}_i \cdot \mathbf{R}_{ij}) (\mathbf{r}_j \cdot 
\mathbf{R}_{ij})}{R_{ij}^5} \right)
\end{aligned}
\end{equation}

We proceed by expanding the resulting Hamiltonian for the two interacting 
excitons in a pair product basis $|{\bf s}_i,{\bf s}_j\rangle=|n_i l_i m_i,n_j 
l_j m_j\rangle$ composed of the single-exciton states $\psi_{n_i,l_i,m_i}({\bf 
r}_i)$ and $\psi_{n_j,l_j,m_j}({\bf r}_j)$, discussed in Sec.~\ref{sec:single}. 
The adiabatic Born-Oppenheimer potentials are then obtained by diagonalizing the 
resulting internal-state Hamiltonian for a given exciton distance $R_{ij}$. Its 
diagonal elements are given by $E_{0,n_il_i}+E_{0,n_jl_j}$ while the 
off-diagonal coupling terms $\langle {\bf s}_i,{\bf s}_j|V^{(ij)}|{\bf 
s}_i^\prime,{\bf s}_j^\prime\rangle$ are calculated using Eq.~(\ref{eq:V_dd}). 
We choose a quantization that is aligned with ${\bf R}_{ij}$, such that the 
total angular momentum $M=m_i+m_j$ is conserved and remains a good quantum 
number for the two-exciton states in the presence of interaction.

The numerical diagonalization then yields potential energy surfaces 
$U_\mu(R_{ij})$ and associated two-exciton states $|\mu({ R}_{ij})\rangle$.
Examples of the resulting interaction curves are shown in 
Fig.~\ref{fig:C6fit}(b) for exciton-pair states around the $15p$ asymptote 
for different values of $M$. The relevant values of $n$, $l$ and $m$ are 
dictated by the band symmetry and the chosen excitation scheme as well as the 
frequency and polarization of the involved excitation lasers. The polarization 
of the laser that drives the Rydberg state transition defines another axis that 
generally can have a finite angle with the chosen quantization axis aligned 
along the distance vector ${\bf R}_{ij}$, such that the optical coupling 
strength can depend on the orientation of the exciton pair through the state 
composition of the two-exciton state $|\mu({ R}_{ij})\rangle$, as discussed 
below.

\section{Excitonic van der Waals interactions} \label{sec:vdW}
The interaction potential and associated two-exciton states assume a simple 
form for large distances $R_{ij}$ where 
\begin{align}
\left|\langle {\bf s}_i,{\bf s}_j|V^{(ij)}|{\bf s}_i^\prime,{\bf 
s}_j^\prime\rangle\right| \!\ll\! 
\left|E_{0,n_il_i}+E_{0,n_jl_j}-E_{0,n_i^\prime l_i^\prime}-E_{0,n_j^\prime 
l_j^\prime}\right|
\end{align}
such that the dipole-dipole interaction only induces a weak far off-resonant 
coupling to other exciton pair states. Due to the aforementioned interaction 
blockade of exciton excitation, this condition can be satisfied in previous 
Cu$_2$O experiments \cite{kazimierczuk2014giant}.
We can thus apply degenerate second-order perturbation theory in the form of an 
effective operator
\begin{equation}
\begin{aligned} \label{eq:vdW}
\hat{H}_{\text{vdW}} &= \left(\frac{e^2}{4\pi\varepsilon_0 \varepsilon_r 
R_{ij}^3 } \right)^2 \sum_{|\alpha \rangle \notin 
\mathcal{M}}\frac{\hat{V}_{11}^{(ij)} |\alpha\rangle \langle \alpha 
|\hat{V}_{11}^{(ij)}}{\delta} \\
 &= \sum_{\mu} \frac{C^{\mu}_{6}}{R_{ij}^6} |\mu\rangle \langle \mu |
\end{aligned}
\end{equation}
whose action is restricted to the degenerate subspaces $\mathcal{M} = \{ | s_i 
s_j\rangle \}$ of fixed $l$ and $M$ at energy $\bar{E}$ \cite{walker2008}. Here 
$\delta = 2\bar{E} - E_{\alpha}$ is the F\"orster defect, while the two-exciton 
eigenstates $|\mu\rangle$ are now independent of the distance $R_{ij}$ but can 
still be composed of several pair states $|{\bf s}_i,{\bf s}_j\rangle$. As shown 
in Fig.~\ref{fig:C6fit}(b) for $n=15$, the van der Waals interaction potential 
obtained in this way provides an excellent description of our numerical results 
already for $R\gtrsim 2.5~\mu$m.

\begin{figure}[ht]
 \begin{center}
 \includegraphics[height=.8\textwidth]{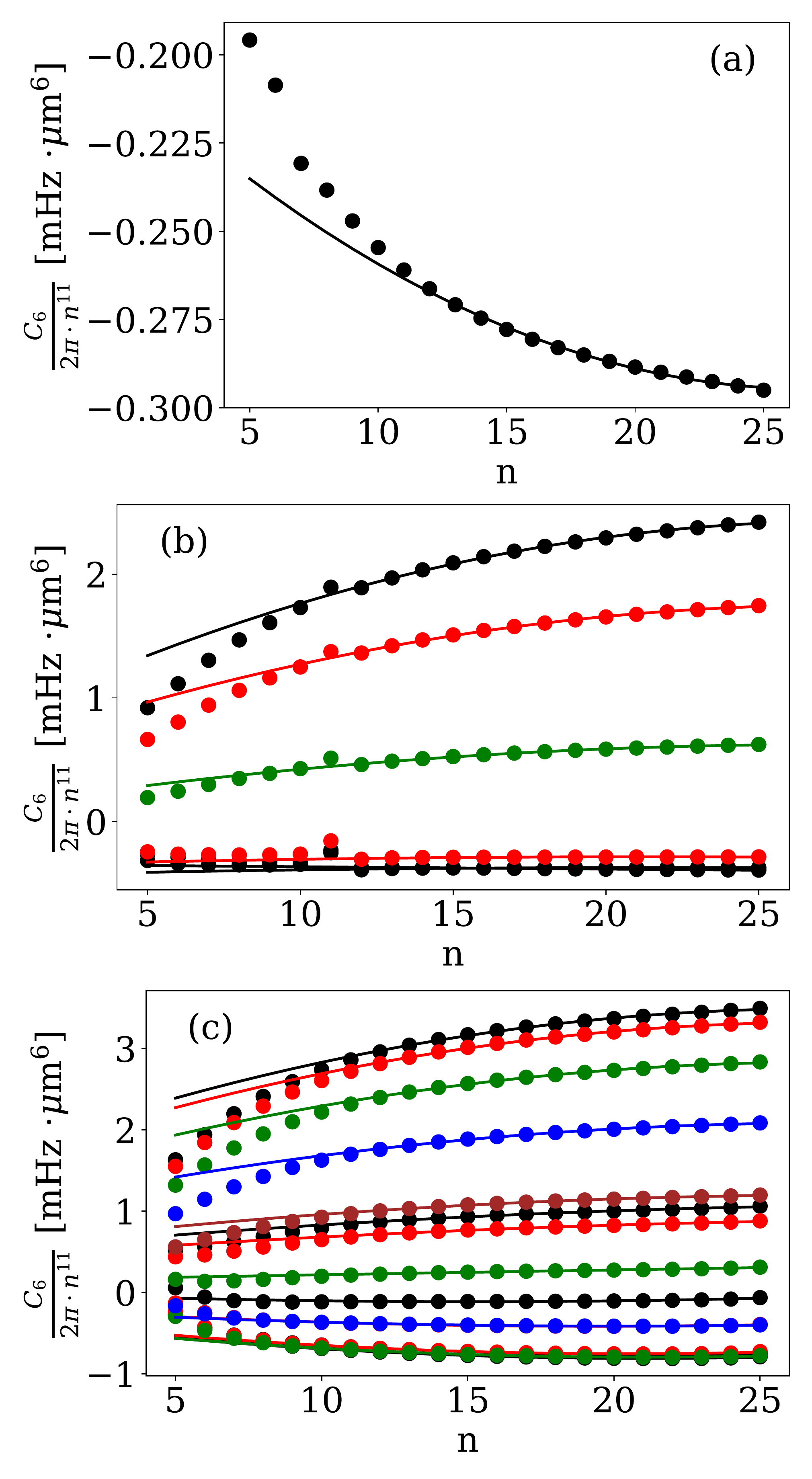}
 \end{center}
 \caption{$C_6$ values for a) the $s-s$, b) $p-p$ and c) $d-d$ asymptotes. The 
fits are for extrapolation for $n \geq 12$. Colored lines denote the families of 
$M$ with $|M|=0$ (black), $|M|=1$ (red), $|M|=2$ (green), $|M|=3$ (blue), 
$|M|=4$ (brown).}\label{fig:C6}
\end{figure}

Figure~\ref{fig:C6} and Tab.~\ref{tab:results} summarize our results for the 
van der Waals interaction between Cu$_2$O Rydberg excitons with angular momenta 
$l=0$ (s), $l=1$ (p) and $l=2$ (d). The simplest asymptote is that of two 
$s$-excitons. With only one asymptotic state  $|n00,n00 \rangle$, 
Eq.~(\ref{eq:vdW}) reduces to standard non-degenerate perturbation theory. For 
higher angular momenta, $l>0$, however, the degenerate pair states get mixed by 
the interaction as given in Tab.~(\ref{tab:results}).
The results are invariant with respect to the sign of $M$, reflecting the 
correponding symmetry of the exciton pair. 

This leaves a total of $2l+1$ different $|M|$-states for a given $l$ and 
$2l+1-|M|$ states within each of the $(l, |M|)$-manifolds, which are indicated 
by different colors in Fig.~\ref{fig:C6fit}(b). The depicted van der Waals 
coefficients and associated eigenstates have been obtained by diagonalizing 
Eq.~(\ref{eq:vdW}) in each $(l, |M|)$-subspace. While the result of this 
calculation may in general depend on the precise value of the principal quantum 
number $n$ through the corresponding coupling strengths to other pair states 
and their relative energy separation, the obtained eigenstates turn out to be
virtually independent of $n$ (cf. standard deviations given in 
Tab.~(\ref{tab:results})). 

\begin{table*}
\begin{center}
\begin{tabular}{|l|l|l|l|l|}
\hline
$\mathbf{|M|}$ & \textbf{composition of $ns-ns$ asymptote} & 
$\frac{\mathbf{c_0}}{2\pi}$ [mHz $\mu$m$^6$] & $\frac{\mathbf{c_1}}{2\pi}$ [mHz 
$\mu$m$^6$] & $\frac{\mathbf{c_2}}{2\pi}$ [mHz $\mu$m$^6$] \\
  &  & $\times 10^{1}$  & $\times 10^{2}$  & $\times 10^{3}$  \\
\hline
\hline
 0 & $\basev{n00}{n00}$ & -2.046  & -0.672  & 0.125  \\
\hline
\hline
$\mathbf{|M|}$ & \textbf{composition of $np-np$ asymptote} &&&\\
2 & $\basev{n11}{n11}$ & 1.257  & 3.641  & -0.666  \\
\hline
1 & $\frac{1}{\sqrt{2}}(\basev{n10}{n11} -\basev{n11}{n10})$ & 5.853  & 8.372  & -1.503 \\
1 & $\frac{1}{\sqrt{2}}(\basev{n10}{n11} +\basev{n11}{n10})$  & -3.574  & 0.680  & -0.160\\
\hline
0 & $(-0.252  \pm  0.003)(\basev{n1-1}{n11}+\basev{n11}{n1-1}) + $ & 8.159  & 11.549  & -2.067\\
& $(0.934  \pm  0.001)\basev{n10}{n10} $  & & & \\
0 & $\frac{1}{\sqrt{2}}(\basev{n1-1}{n11} - \basev{n11}{n1-1})$ & -3.456  & -0.205  & 
0.006\\
0 & $(0.661  \pm  0.001)(\basev{n1-1}{n11}+\basev{n11}{n1-1}) +$ & -4.371  & 0.608  & -0.143 \\
& $(0.356  \pm  0.004)\basev{n10}{n10}$ &&& \\
\hline
\hline
$\mathbf{|M|}$ & \textbf{composition of $nd-nd$ asymptote} &  &  &  \\
4 & $\basev{n22}{n22}$ & 6.247  & 4.067  & -0.719 \\
\hline
3 & $\frac{1}{\sqrt{2}}(\basev{n21}{n22} +\basev{n22}{n21})$ & -2.201  & -1.936  & 0.481\\
3 & $\frac{1}{\sqrt{2}}(\basev{n21}{n22} -\basev{n22}{n21})$ & 10.906  & 7.237  & -1.317\\
\hline
2 & $-0.429(\basev{n20}{n22}+\basev{n22}{n20})+0.795\basev{n21}{n21}$ & 14.881  & 9.862  & -1.807 \\
2 & $\frac{1}{\sqrt{2}}(\basev{n20}{n22}-\basev{n22}{n20})$ & 1.536  & 0.650  & -0.018\\
2 & $0.562(\basev{n20}{n22}+\basev{n22}{n20})+0.607\basev{n21}{n21}$ & -4.005  & -3.752  & 0.889 \\
\hline
1 & $0.218(\basev{n2-1}{n22}-\basev{n22}{n2-1})-0.673(\basev{n20}{n21}-\basev{n21}{n20})$ & 17.480  & 
11.564  & -2.125 \\
1 & $(-0.465 \pm 0.001)(\basev{n2-1}{n22}+\basev{n22}{n2-1})+ $ & 4.604  & 2.582  & -0.373\\
& $(0.533 \pm 0.001)(\basev{n20}{n21}+\basev{n21}{n20})$ &&& \\
1 & $(0.533\pm0.001)(\basev{n2-1}{n22}+\basev{n22}{n2-1})+$ & -3.559  & -3.905  & 0.954\\
& $(0.465\pm0.001)(\basev{n20}{n21}+\basev{n21}{n20})$ & && \\
1 & $-0.673(\basev{n2-1}{n22}-\basev{n22}{n2-1})-0.218(\basev{n20}{n21}-\basev{n21}{n20})$ & -2.017  & -2.208  & 0.567 \\
\hline
0 & $-0.082(\basev{n2-2}{n22}+\basev{n22}{n2-2})+$ & 18.386  & 12.153  & -2.234\\
& $0.451(\basev{n2-1}{n21}+\basev{n21}{n2-1})-0.762\basev{n20}{n20}$ &&& \\
0 & $(-0.222  \pm  0.001)(\basev{n2-2}{n22}-\basev{n22}{n2-2})+$ & 5.559  & 3.240  & -0.501\\
& $0.671(\basev{n2-1}{n21}-\basev{n21}{n2-1})$ &&& \\
0 & $(0.343  \pm  0.004)(\basev{n2-2}{n22} + \basev{n22}{n2-2}) + $ & -0.146  & -1.315  & 0.434 \\
 & $(-0.448  \pm  0.003)(\basev{n2-1}{n21}+\basev{n21}{n2-1}) + $&&&\\
 & $(-0.603  \pm  0.001)\basev{n20}{n20}$ &  & & \\
0 & $(0.613  \pm  0.002)(\basev{n2-2}{n22}+\basev{n22}{n2-2}) + $ & -3.861  & -4.031  & 0.956\\
 & $(0.311  \pm  0.003)(\basev{n2-1}{n21}+\basev{n21}{n2-1})+$ &&&\\
 &$(0.236  \pm  0.004)\basev{n20}{n20}$ & & & \\
0 & $-0.671(\basev{n2-2}{n22}-\basev{n22}{n2-2})+$ &&&\\
& $(-0.222  \pm  0.001)(\basev{n2-1}{n21}-\basev{n21}{n2-1})$ & -3.736  & -3.639  & 0.850\\
\hline
 \end{tabular}
 \end{center}
\caption{Various asymptotes listed by quantum numbers $l, M$ with corresponding 
approximate asymptotic wavefunctions and van der Waals coefficients $C_6(n) = 
n^{11} (c_0 + c_1 n + c_2 n^2)$ as obtained from fitting for principal quantum 
numbers $n=12\text{--}25$. The given errors are standard deviations calculated 
from the numerically obtained eigenfunctions. }\label{tab:results}
\end{table*}

The van der Waals interaction rapidly increases with the principal quantum 
number $n$. This is due to the quadratic increase of the dipole matrix elements 
for transitions between Rydberg states and the decreasing level spacing, such 
that $\delta \sim n^{-3}$, which overall results in an increase of the van der 
Waals coefficient as $C_6\sim n^{11}$. Similar to the behavior of atomic systems 
\cite{singer2005}, our numerical results can be well described by the slightly 
modified scaling relation
\begin{align}
 C_6(n) = n^{11}\left( c_0 + c_1 n^1 + c_2 n^2 \right),
\end{align}
whose coefficients $c_i$ depend on the angular numbers and are given in 
Tab.~\ref{tab:results}. As shown in Fig.~\ref{fig:C6}, this simple expression 
permits an accurate determination of the van der Waals interaction for the 
depicted range $12\lesssim n\lesssim 25$.

 \begin{figure}[ht]
 \begin{center}
 \includegraphics[height=.18\textwidth]{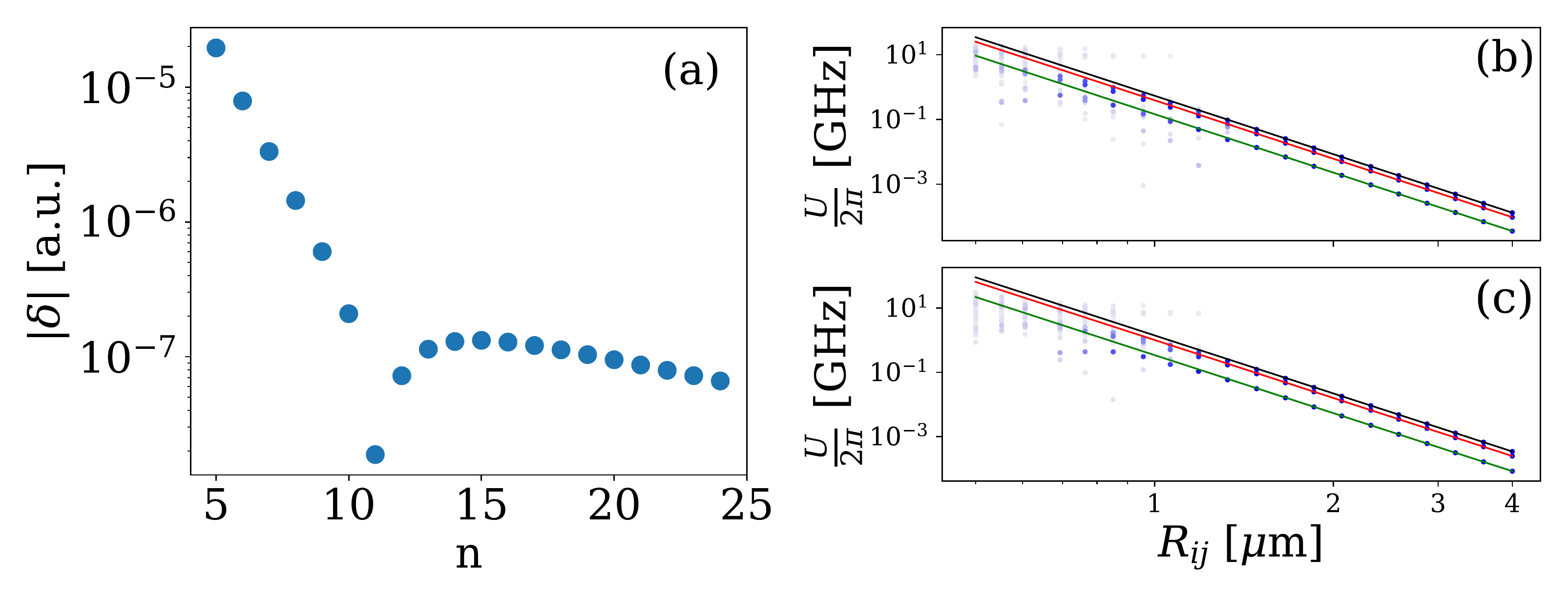}
 \end{center}
 \caption{At $n=11$ a near F\"orster resonance in the channel $np + np 
\rightarrow (n-1)d + (n+1)d$ leads to outlying points in the otherwise quite 
homogeneous range of $C_6(n)$. b) The F\"{o}rster defect $\delta$ of the given 
channel passes zero near $n=11$. Taking the repulsive part of the interaction as 
an example, the numerical solution (blue dots) is compared with the long-range 
asymptote with $|M|=0$ (black), $|M|=1$ (red), $|M|=2$ (green) for $n=11$ (b) 
and $n=12$ (c). The color code denotes the relative $p$-component of each 
asymptote. Despite of the near resonance at $n=11$, their difference is small as 
long as $R \gtrsim 1\mu$m since the dominant channels fall into the van der 
Waals regime.}
\label{fig:resonance}
\end{figure}

The van der Waals coefficients of the $np-np$ asymptotes for $n=11$, however, 
show slightly larger deviations. This is due to the $np + np \rightarrow (n-1)d 
+ (n+1)d$ coupling channel, which becomes near resonant around $n=11$ as shown 
in Fig.~\ref{fig:resonance}(a). As the denominator in Eq.~(\ref{eq:vdW}) goes 
through a minimum, the resulting van der Waals interaction is enhanced, while 
the validity of Eq.~(\ref{eq:vdW}) requires larger exciton distances. However, a 
comparison with our numerical results [Fig. (\ref{fig:resonance}(b)] shows that 
the agreement remains good even at relatively small exciton separations of 
$\sim1\mu$m, comparable to what is also required for $n=12$ in the absence of 
the F\"orster resonance. 

\begin{figure}[ht]
 \begin{center}
 \includegraphics[height=.3\textwidth]{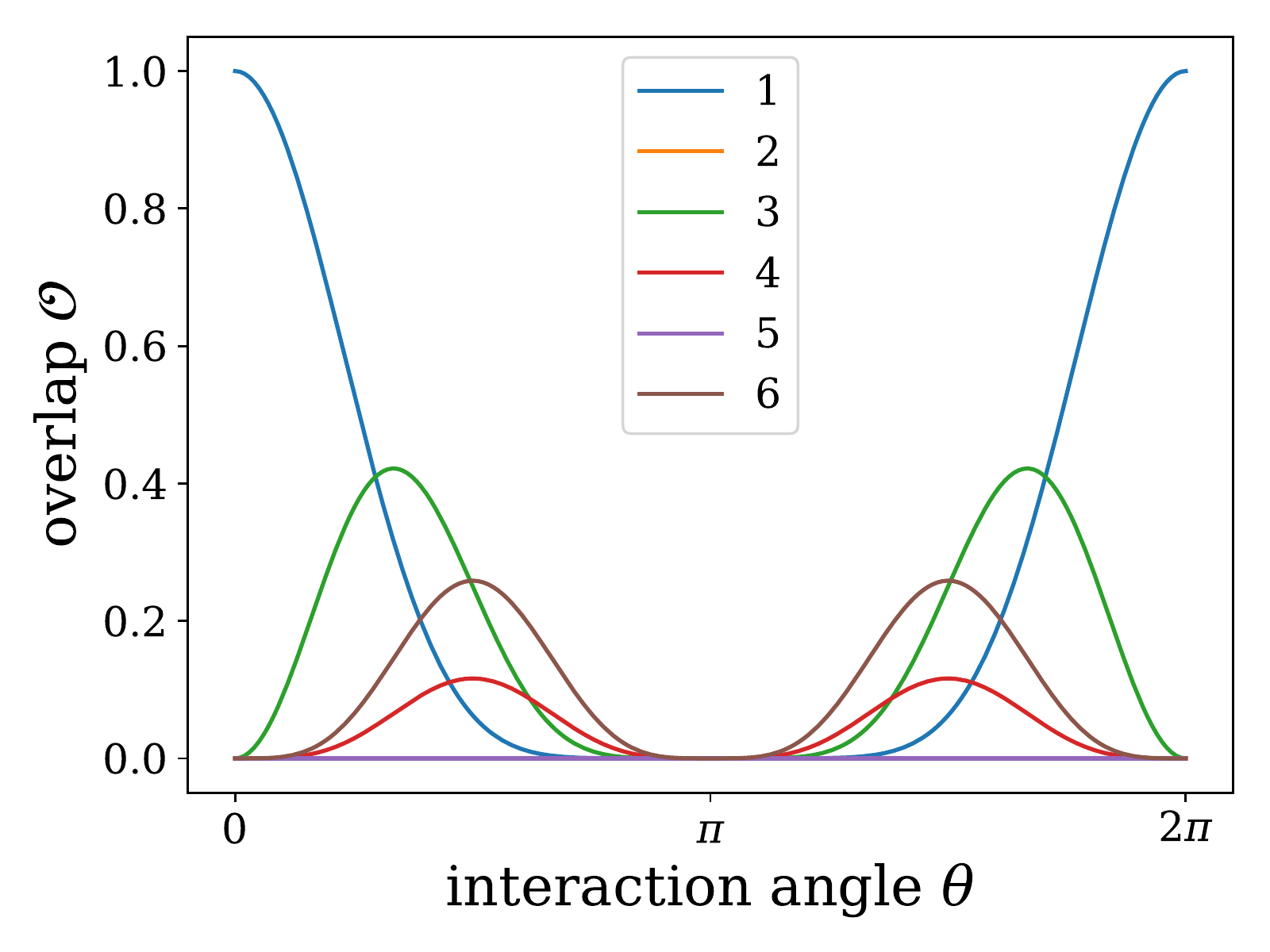}
 \end{center}
\caption{Overlap of the optically active pair state $\phi_a$ with the 
asymptotic molecular eigenstates $|\mu\rangle$, $\mathcal{O} = |\langle \phi_a 
| \mu \rangle|^2$. Here, for illustration, we chose $|\phi_a\rangle = 
|n11,n11\rangle$ for $\sigma^+$-light and the eigenstates of the 
$p-p$-asymptote (labeled as listed in Tab. \ref{tab:results}). Note that 
antisymmetric states 2 and 5 do not couple to the excitation laser.}
\label{fig:rotation}
\end{figure}

An interesting and often relevant situation arises when an external field 
introduces an axis that is not parallel to the intermolecular axis 
$\mathbf{R}_{ij}$. Examples include electric and magnetic fields as well as a 
tilted excitation laser, each defining a new axis $\hat{\mathbf{z}}^\text{lab}$. 
Without loss of generality, we assume that the molecular axis lies in the 
$(x,z)$-plane of the laboratory frame, such that the two 
$\hat{\mathbf{z}}$-axes span the interaction angle $\theta$ \cite{weber2016}. A 
general transformation of the states between the frames is given by
\begin{align}
|n l m\rangle^\text{mol} = \sum_{m^\prime} \left[d_{m 
m^\prime}^{l}(\theta)\right]^* |n l m^\prime \rangle^\text{lab},
\end{align}
where $d_{m m^\prime}^{l}(\theta)$ denotes elements of the lowercase Wigner 
d-matrix \cite{wigner1959}. While it is often advantageous to express the 
external field in the molecular frame, we illustrate the angular dependence by 
evaluating the optical coupling strengths of the $p-p$ asymptotic pair states 
in the laboratory frame. For a definite laser polarization, only certain pair 
states are optically active and the optical coupling, given by the their overlap 
with the optically active pair state, becomes a function of $\theta$ 
(Fig.~\ref{fig:rotation}). 

\section{Discussion} \label{sec:discussion}

In summary, we have evaluated the interaction between Rydberg excitons in 
Cu$_2$O semiconductors and provided an expression that, together with the 
tabulated parameters, facilitates a simple and yet accurate determination of 
the resulting van der Waals interaction for a broad range of Rydberg states. 
Such van der Waals interactions may be responsible for the recently observed 
\cite{kazimierczuk2014giant} excitation blockade of excitons in Cu$_2$O. The 
highest lying exciton state reported in these experiments ($n=25$) covers a 4 
million times larger volume than the $2p$-exciton state, owing to the $\sim n^2$ 
scaling of the exciton radius. Such a large radius entails an even higher 
enhancement of the  polarizability as $\sim n^7$, such that electrostatic 
interactions become relevant at exciton separations where exchange effects are 
negligible. 

The importance of long-range dipole interactions for Rydberg excitons is 
connected to the way they are created by optical excitation.  
Shifts of the Rydberg pair-state energy due to exciton-exciton interactions can 
inhibit the simultaneous generation of Rydberg excitons within a certain radius 
once they exceed the width of the corresponding exciton line. For strong 
interactions and sufficiently narrow excitation lines, this excitation blockade 
effect thus ensures that excitons are only created at distances where van der 
Waals interactions dominate. This may open up a new regime where strong 
interaction effects become observable at very low densities of excitons, which 
therefore interact over long distances in a quasi-static fashion, as opposed to 
short-range collisional interactions that determine the behaviour of 
ground-state excitons.
The accurate knowledge of van der Waals interactions between Rydberg excitons, 
as provided by the present work, enables quantitative theoretical studies of this
blockade effect. This in turn would also make it possible to estimate the  
importance of other mechanisms such as interactions with the free charges of 
potentially forming electron-hole plasmas \cite{heckoetter2017} and to thereby determine their 
relative contribution to the nonlinear optical response of the semiconductor.

One major difference between the typical scales of Rydberg states of excitons 
and atomic Rydberg states stems from the effective electron and hole masses as 
well as the dielectric constant, $\varepsilon_r$, of the semiconductor. Both 
factors tend to decrease the binding energy and lead to a decrease of the 
Rydberg constant by a factor $\nu = \mu_X/(\mu_A\varepsilon_r^2)$, where $\mu_X$ and 
$\mu_A$ denote the reduced mass of the excitonic and atomic system, 
respectively. On the other hand, the excitonic radius is increased by a factor
$(\varepsilon_r \nu)^{-1}$.
Therefore we expect the van der Waals coefficient to 
increase as  $\sim \varepsilon_r^4\mu_A^5/\mu_X^5$. Accordingly, the van der 
Waals coefficients as calculated in the present work exceed those of typical 
atomic Rydberg states with comparable quantum numbers 
\cite{singer2005,mukherjee2011} by 5 orders of magnitude.
This also opens the search for other suitable semiconductor systems with Rydberg states \cite{chernikov2014exciton, Greenland2010}, each featuring different interaction properties and additional rich physics \cite{wang2018}.

Rydberg excitons thus suggest promising avenues to studies of strong 
interaction effects in confined geometries \cite{krueger2018}, optical 
nonlinearities \cite{walther2018giant} or nonclassical light generation 
\cite{bloch2018, khanzali2017} at ultralow exciton densities. The results of the present work 
provide simple yet accurate interaction potentials for future theoretical 
explorations of these perspectives.

\section{Acknowledgement}
We would like to thank the authors of Ref.~\cite{weber2016} for sharing their 
Rydberg potential software ``pairinteraction'', parts of which we used in the 
exact diagonalization. We are grateful to the DFG SPP 1929 GiRyd for financial 
support.


%

\end{document}